\documentclass{PoS}

\newcommand{\eq}[1]{(\ref{#1})}

\title{
\vspace{-1cm}
\begin{minipage}{\textwidth}
\begin{flushright}
\texttt{\footnotesize
ITEP-LAT-2009-15\\
}
\end{flushright}
\end{minipage}\\[15pt]
Gluon propagators and center vortices\\
at finite temperature}

\ShortTitle{Gluon propagators and center vortices}

\author{\speaker{T. Saito}\\ 
       Integrated Information Center, Kochi University, Akebono-cho,
       Kochi, 780-8520, Japan\\
       E-mail: \email{tsaito@rcnp.osaka-u.ac.jp}}
\author{M.~N.~Chernodub\\
        CNRS, Laboratoire de Mathematiques et Physique Theorique,
        F\'ed\'eration Denis Poisson, Universit\'e de Tours,
        Parc de Grandmont, F37200, Tours, France \\
        Department of Mathematical Physics and Astronomy, University of Gent,
        Krijgslaan 281, S9, Gent, B-9000 Belgium \\
        Institute of Theoretical and Experimental Physics ITEP, 117259 Moscow, Russia\\
        E-mail: \email{Maxim.Chernodub@lmpt.univ-tours.fr}}
\author{Atsushi Nakamura\\
        Research Institute for Information Science and Education,
        Hiroshima University, Higashi-Hiroshima, 739-8527, Japan\\
        E-mail: \email{nakamura@riise.hiroshima-u.ac.jp}}
\author{V.~I.~Zakharov\\
        Institute of Theoretical and Experimental Physics ITEP, 117259
        Moscow, Russia\\
        Max-Planck Institut f\"ur Physik, F\"ohringer Ring 6, 80805,
        M\"unchen, Germany\\
        E-mail: \email{xxz@mppmu.mpg.de}}

\abstract{We study influence of center vortices on infrared properties of gluons in the deconfinement phase of quenched QCD.
We observe a significant suppression of the magnetic component of the gluon propagator in the low-momentum region after the vortices
are removed from the gluon configurations. The propagator of the electric gluon stays almost unaffected by the vortex removal.
Our results demonstrate that the center vortices are responsible for important nonperturbative properties of the
magnetic component of the quark-gluon plasma.}

\FullConference{The XXVII International Symposium on Lattice Field Theory - LAT2009\\
         July 26-31 2009\\
         Peking University, Beijing, China}

\begin{document}

\section{Introduction}

The quark-gluon plasma attracts a lot of attention nowadays due to a
real feasibility to create this unusual state of matter in heavy-ion
collisions. The wide interest to this topic is additionally heated
by discoveries of interesting properties of the plasma (the very low
viscosity of this substance is a well-known
example~\cite{viscosity}). Moreover, it seems plausible that the
quark-gluon plasma may contain exotic objects such as magnetic
monopoles~\cite{ref:monopoles,ref:Shuryak} and magnetic
vortices~\cite{ref:monopoles}. Both the monopoles and the vortices
constitute the nonperturbative magnetic component of the plasma. In
this paper we concentrate ourselves on the magnetic vortices. Our
aim is to demonstrate from the first principles that these objects
are crucially important for the properties of the quark gluon plasma,
because the vortices affect long-range propagation of hot magnetic gluons.

Originally, the magnetic vortices were invoked to explain quark confinement in QCD (for a review see Ref.~\cite{ref:greensite}).
The magnetic vortices are certain stringlike
configuration of gluons which populate the vacuum of non-Abelian gauge theories. According to the vortex picture, the confining
force between colored objects emerges due to spatial percolation of the magnetic vortex strings because the vortices
lead to certain amount of disorder. The value of the Wilson loop
changes by a center element of the gauge group if the magnetic vortex pierces the loop~\footnote{The ``magnetic vortices'' and
the ``center vortices'' are the same objects in our terminology.}. Therefore, very large loops receive
rapidly fluctuating contributions
from the vortex ensembles. These fluctuations make the average value of the Wilson loop very small. One can show that the suppression
of the loop follows an area law for very large loops implying a linear confining potential between a static quark and
an antiquark~\cite{ref:greensite}.

The relevance of the vortices to the confining properties of the vacuum can be demonstrated using an elegant method of vortex removal~\cite{ref:massimo}. The gluons in the non-Abelian field configurations can be divided into the two parts:
``gluons emerging due to the magnetic vortices'' and ``the rest''. It turns out that if the vortices are removed, the remaining
gluons cannot support the confining force between the quarks and antiquarks, and the confinement is lost at low temperatures~\cite{ref:massimo}.
On the other hand, the long-distance confinement is unaffected by an inverse procedure which keeps the vortices intact
and removes ``the rest'' from the gauge field configurations. Thus, the magnetic
vortices carry crucial information about the quark confinement at low temperatures.

It was suggested in Ref.~\cite{ref:monopoles} that in the deconfinement (quark-gluon) phase the vortices become {\it real} objects,
as they form a magnetic component of the thermal plasma similarly to the thermal monopoles. The vortices become real because
they provide a large contribution to the thermodynamics of the system according to the numerical simulations of Ref.~~\cite{ref:thermodynamics}.

Another known important property of the vortices is their role in the propagation of gluons at zero temperature~\cite{ref:gluons:vortices}.
The presence of the vortices in the gluon ensembles enhance the gluon propagators in the low-momentum region.
As the vortices do not affect the high-momentum the gluons, one concludes that the vortices are responsible for the long-distance gluon propagation.

In this paper we investigate the propagation of the gluons in the high-temperature gluon plasma phase, which is interesting
from the point of view of the heavy-ion collisions at RHIC and LHC. The gluon propagators in the deconfinement phase were studied
in details in Ref.~\cite{NSS}. In this paper we combine the methods of Ref.~\cite{ref:gluons:vortices} and Ref.~\cite{NSS} to
find the effect of the magnetic vortices on the propagation of the hot gluons. We found that the effect is not trivial.

\section{Propagation of hot gluons}

\subsection{Gluon propagators}

We define the gauge potential $A$ via the lattice link variable $U$ in a standard way,
\begin{equation}
A_{\mu}^a(x,t) = \frac{1}{2} \mbox{Tr} \, \sigma^a U_{\mu}(x,t)\,.
\end{equation}
The gluon correlation function is given by the formula:
\begin{equation}
D_{\mu\nu}(x,t) = \left\langle A_{\mu}(0,0) A_{\nu}^{*}(x,t) \right\rangle\,.
\label{eq:D:space}
\end{equation}
We are interested in the momentum-space propagators which are given by a Fourier transform of \eq{eq:D:space} evaluated either in the
Coulomb ($\partial_i A_i = 0$) or Landau ($\partial_\mu A_\mu = 0$) gauges.

We study the propagation of the gluons with the zero energy transfer, $q_0 = 0$,
\begin{equation}
D_{\mu\nu}(\vec{q},q_0=0) = \frac{1}{N_t} \sum_t D_{\mu\nu}(\vec{q},t)\,,
\label{eq:Dmunu}
\end{equation}
where $N_t$ is the temporal extent of the lattice.

At finite temperature the temporal component of (\ref{eq:Dmunu}) corresponds to
the electric gluon, $D_E \sim D_{00}$, while the propagator of the magnetic gluon
is defined by the spatial correlations, $D_M \sim D_{ii}$.
The masses corresponding to the electric and magnetic gluons
can conveniently be calculated with the help of, respectively, $D_E$ and $D_M$ propagators
in the coordinate space~\eq{eq:D:space} at large spatial separations~\cite{NSS}.
Here we study the momentum dependence of both electric and magnetic gluon propagators
in order to investigate the effect of the vortices both in the infrared (large distance) and
ultraviolet (short distance) regions.

\subsection{Gluons at finite temperature}

At finite temperature the electric and magnetic gluons behave differently.
An electric gluon demonstrates the color-screening of the Debye type. The corresponding
potential is proportional to $\exp(-m_e R)/R$, where $R$ is the spatial separation.
According to the perturbation theory the temperature dependence of the electric mass is
$m_e \sim g(T) T$, where $g(T)$ is the running coupling at the temperature scale $T$.

It is hard to define the magnetic gluon propagator in the perturbative theory since
the magnetic sector of QCD has a nonperturbative nature. The magnetic mass is an
important quantity because it serves as an infrared cutoff of a thermal QCD theory.
In the scope of the dimensional reduction the magnetic mass should scale as $m_m \sim g^2(T)T$ at $T \gg T_c$.
This temperature dependence agrees with the recent lattice simulations in the temperature
range $T/T_c=1.5 \sim 6$, Ref.~\cite{NSS}.
The temperature behavior of another nonperturbative quantity, the spatial string tension, is also understood in terms of
the magnetic scaling~\cite{Bali}, $\sigma_{\mathrm{sp}} \sim [g^2(T)T]^2$.
Thus, the magnetic gluons play an important role in the infrared physics of the deconfinement phase.

\subsection{Magnetic vortices}

In order to identify the center vortex on the lattice we employ the direct maximal center gauge (MCG)~\cite{DFGGO}
in SU(2) gauge theory. The gauge condition is given by the maximization of the functional $R[U]$ over all possible
gauge transformations $U \to U^\Omega$,
\begin{equation}
\max_{\Omega} R[U^\Omega]\,, \qquad \qquad R[U] = \frac{1}{VT} \sum_{x,t} \mbox{Tr} \, \left[ U_{\mu}(x,t) \right]^2\,.
\label{mcpe}
\end{equation}
The center gauge field is defined as the $\mathbb{Z}_2$-valued link field $Z_{\mu}(x) = \mbox{sgn} \mbox{Tr} \left[ U_{\mu}(x) \right] = \pm 1$
in the MCG. If a $\mathbb{Z}_2$ plaquette $Z_{x,\mu\nu} = Z_\mu(x) Z_\nu(x+\hat\mu) Z_\mu(x+\hat\nu) Z_\nu(x)$ takes a negative value,
then this plaquette is pierced by a center vortex. A detailed review of the magnetic (center) vortices can be found in Ref.~\cite{ref:greensite}.

The vortices can be removed using the de Forcrand-D'Elia procedure~\cite{ref:massimo} which is formulated as the redefinition
of the original gauge field
\begin{equation}
U_{\mu}(x) \to U_{\mu}^{'}(x) = Z_{\mu} \, U_{\mu}(x)\,.
\label{eq:massimo}
\end{equation}
The gauge field $U'$ does not contain vortices.
Numerical simulations show that the string tension vanishes in the confinement phase after the removal of the center vortices
\cite{ref:massimo}.
The infrared propagation of the cold gluons is also suppressed by the vortex removal~\cite{ref:gluons:vortices}. Below we study the effect
of the vortices on the propagation of the hot gluons using the procedure~\eq{eq:massimo}.

\section{Numerical Results}

We simulate the $SU(2)$ lattice theory in the quenched approximation. We generate $10 \sim 30$ gauge configurations per a fixed set
of lattice couplings. We use several lattice volumes $N_s^3 N_t$
with varying spatial size $N_s=\{12,20,24,32,48\}$ and fixed temporal size $N_t=4$.
We perform the MCG and then Landau (or Coulomb) gauge fixing. The convergence criteria
for the violation of the gauge fixing conditions
are set as $\epsilon = 10^{-10}$ and $\epsilon = 10^{-8}$,
respectively.
We work at two temperatures: close to the transition, $T \approx 1.40\, T_c$, and deeply in the deconfinement phase, $T \approx 6.0 T_c$
(the later regime may be realized at LHC/ALICE experiments).

\subsection{Electric and magnetic gluons in Landau and Coulomb gauges}

\begin{figure}[htbp]
\includegraphics[scale=0.29]{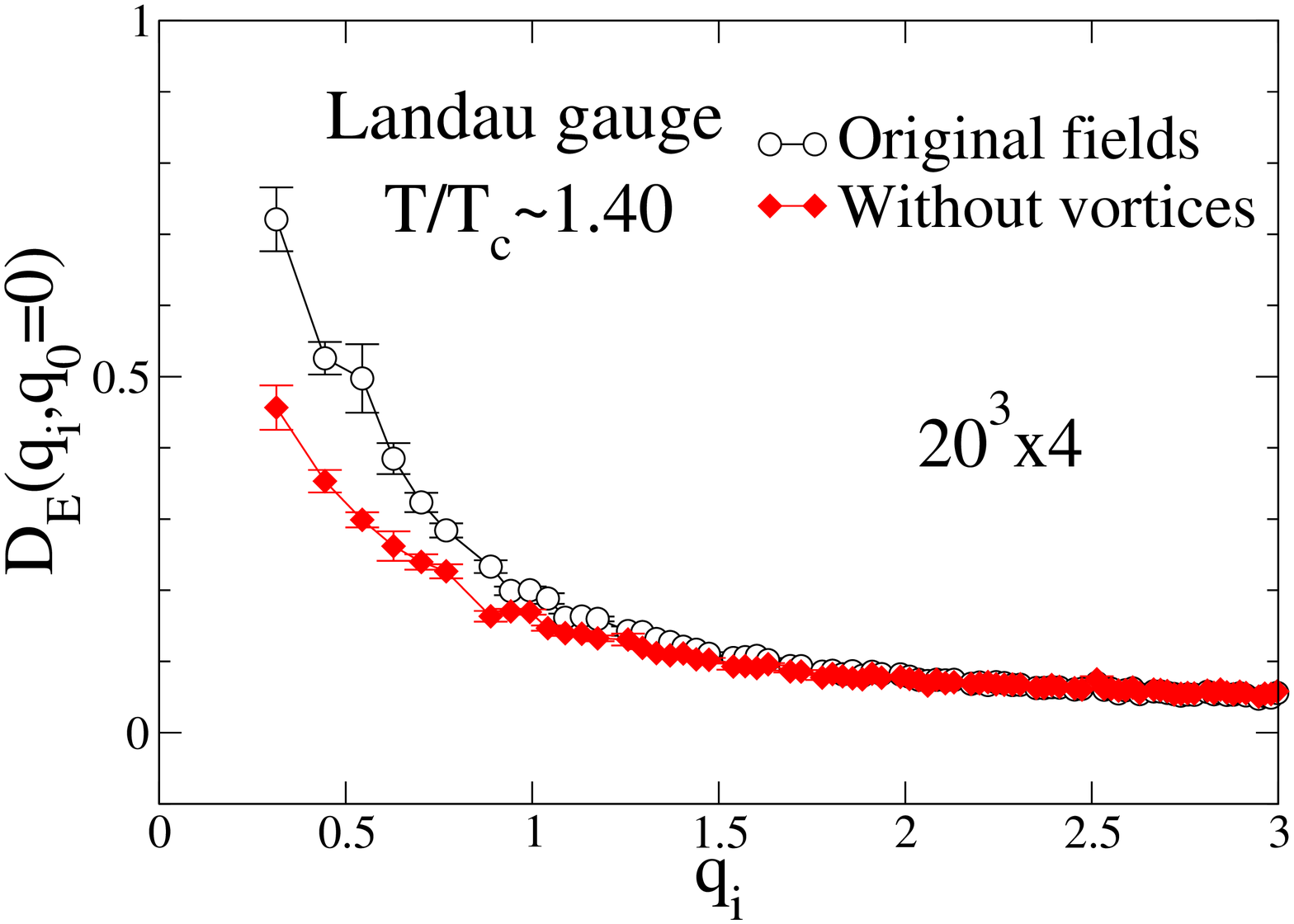}
\hskip -9mm
\includegraphics[scale=0.29]{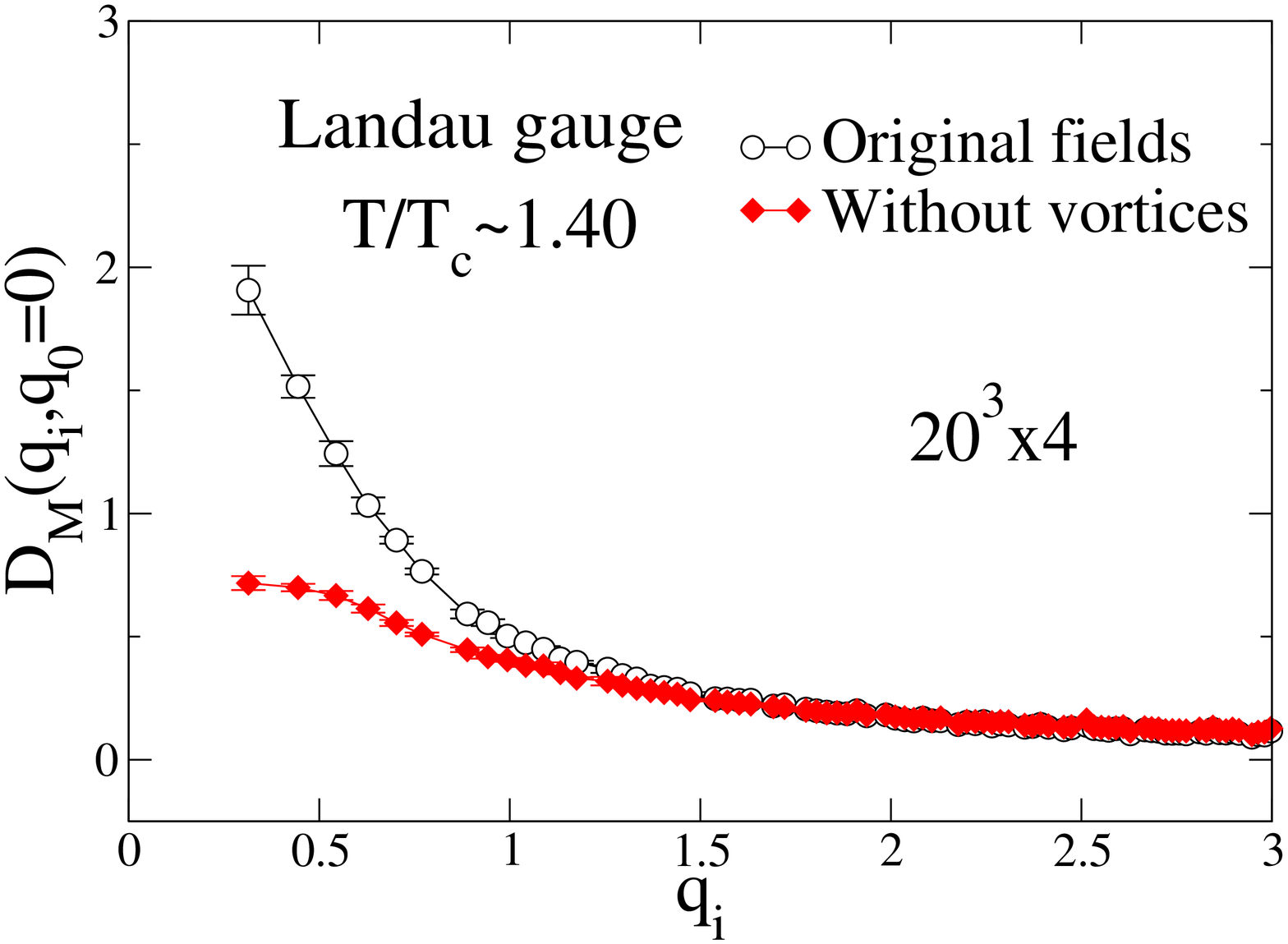}
\caption{The electric (left) and magnetic (right) gluon propagators as functions of the spatial momentum. The calculations are done
in the Landau gauge using original and vortex removed gauge field configurations.}
\label{fig1}
\end{figure}

In Fig.~\ref{fig1} we demonstrate the electric and magnetic gluon propagators at $T = 1.40\, T_c$ calculated both at the original
gluon configurations $U$ and at the gluon configurations without vortices, $U'$, Eq.~\eq{eq:massimo}.
One immediately notices that the vortices do not affect the high momentum region of both electric and magnetic propagators.
This feature -- which is in line with the zero-temperature studies of Ref.~\cite{ref:gluons:vortices} -- is a natural consequence of the fact that the center vortices are nonperturbative objects so that they may not play a role in the perturbative regime.

However, the magnetic vortices affect the long-distance propagation of the gluons. After the removal of the vortices the magnetic
propagator gets suppressed in the infrared region, Fig.~\ref{fig1}(right). The suppression of the infrared electric gluons is also
visible [Fig.~\ref{fig1}(left)], but this effect is much less compared to the suppression in the magnetic sector.

Thus, the center vortices are related to the magnetic sector of quark-gluon plasma phase
as it was suggested in Ref.~\cite{ref:monopoles}. The vortices support the long-distance propagation of the magnetic gluons
and they may be responsible for the low viscosity of the QCD plasma~\cite{ref:monopoles}.

\begin{figure}[htbp]
\includegraphics[scale=0.28]{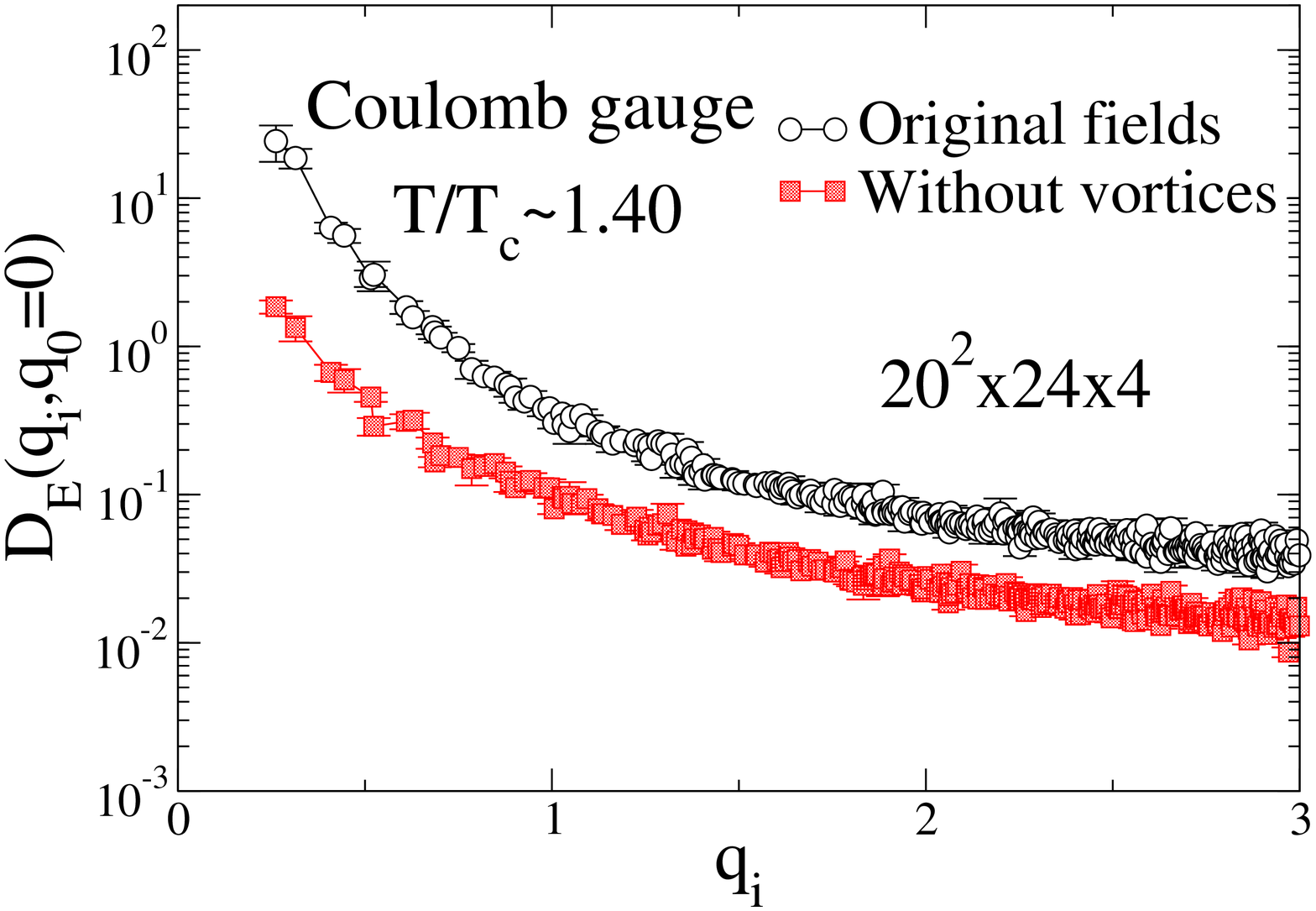}
\hskip -6mm
\includegraphics[scale=0.28]{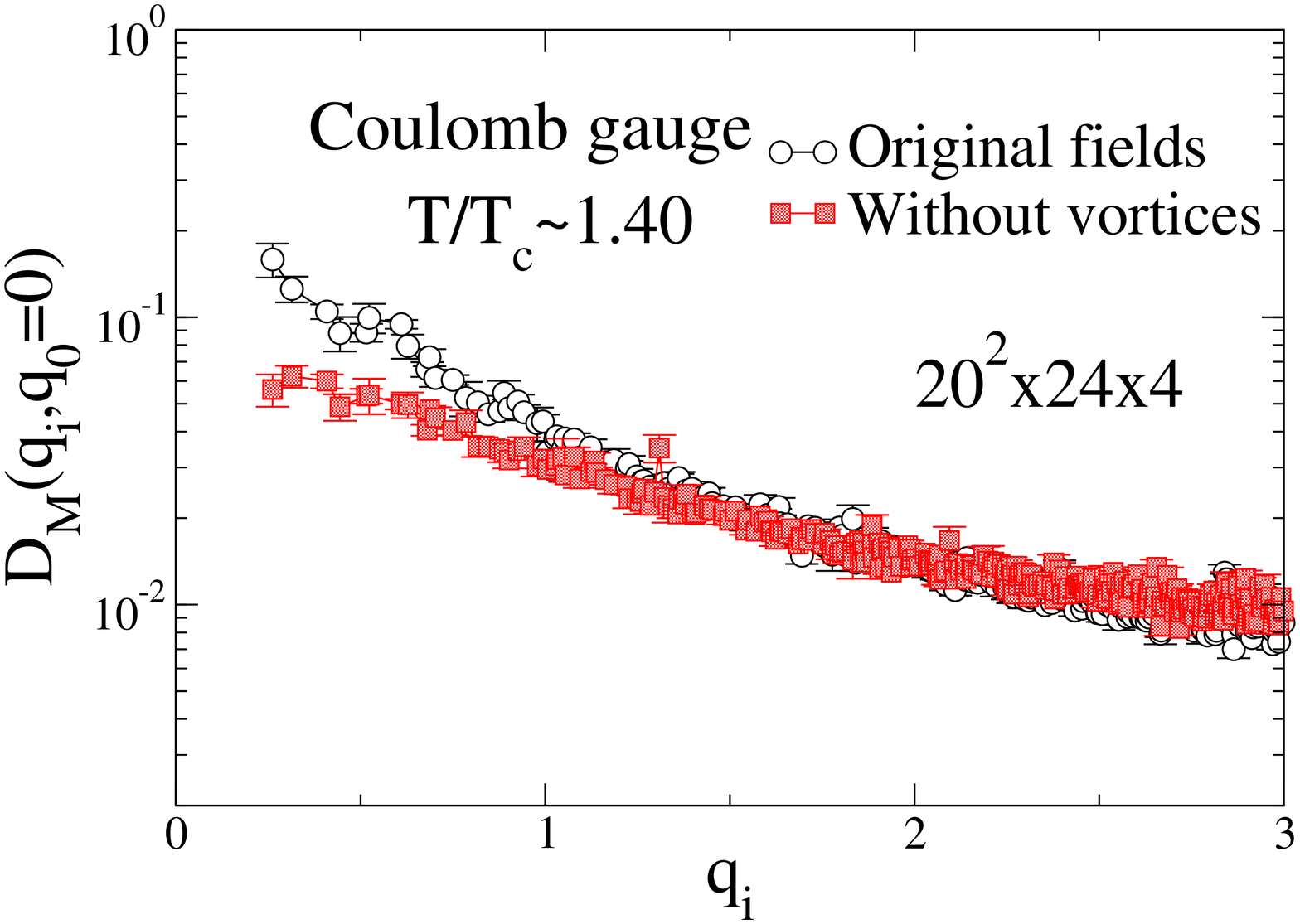}
\caption{The same as in Fig.~1 but in the Coulomb gauge. Notice the logarithmic scale of the ordinate axis.}
\label{fig2}
\end{figure}
We also studied the propagation of the hot gluons in the Coulomb gauge\footnote{Note that our gauge field configurations
are additionally fixed by the temporal gauge fixing that maintain the Coulomb gauge property (for the details one can consult Ref.~\cite{NSS}).}.
Our numerical results are shown in Fig.~\ref{fig2}. It is interesting to note that the removal of the vortices has a renormalization-like effect
on the electric gluons. The electric propagator is scaled by a constant factor which is visibly independent on the momentum.
The propagator of the magnetic gluon in the Coulomb gauge
is suppressed in the infrared region similarly to the magnetic propagator in the Landau gauge. Thus,
the effect of the infrared suppression is common for the Landau and Coulomb gauges. Below we continue the discussion of the gluon
propagators concentrating only on the Landau gauge.

\begin{figure}[htbp]
\includegraphics[scale=0.28]{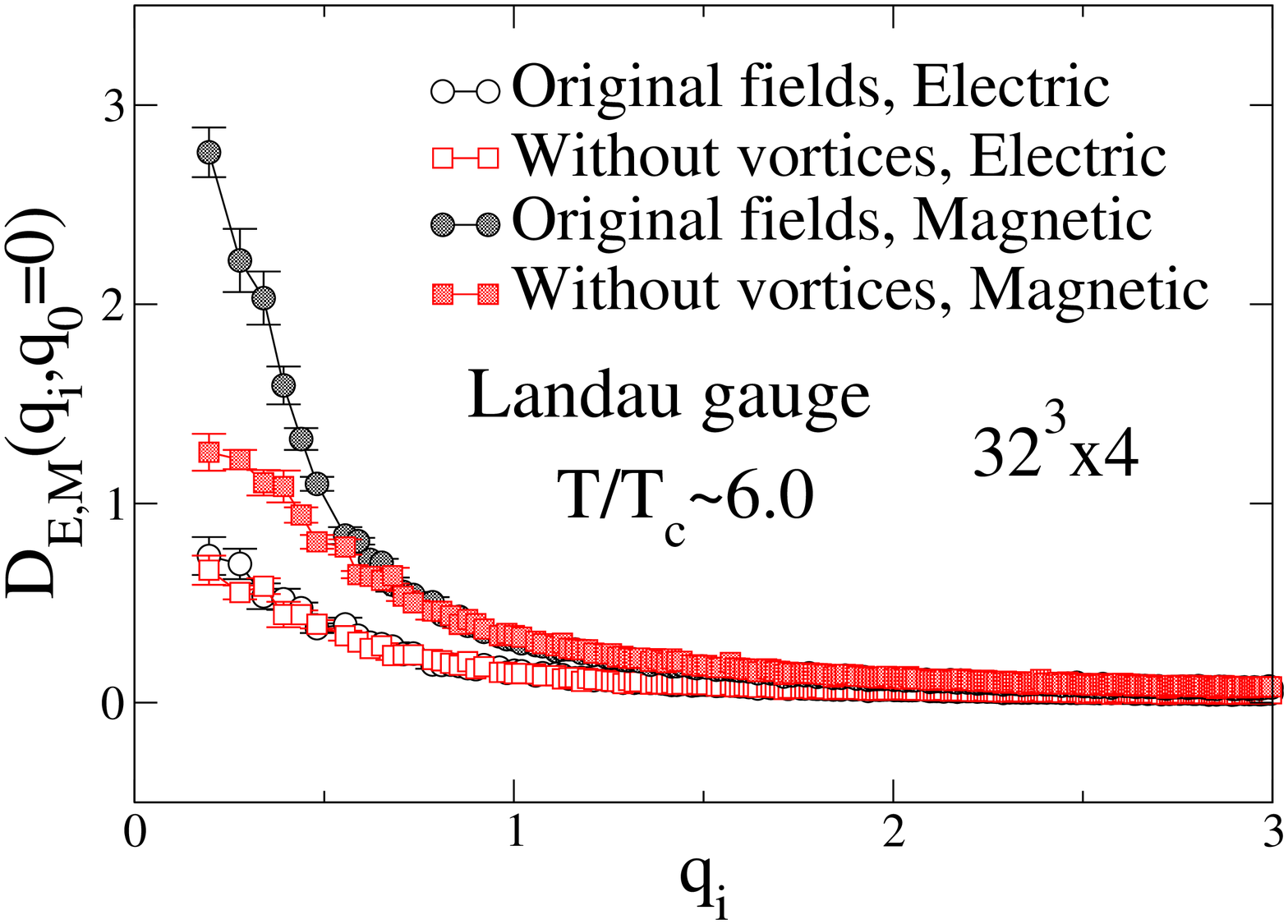}
\hskip -6mm
\includegraphics[scale=0.28]{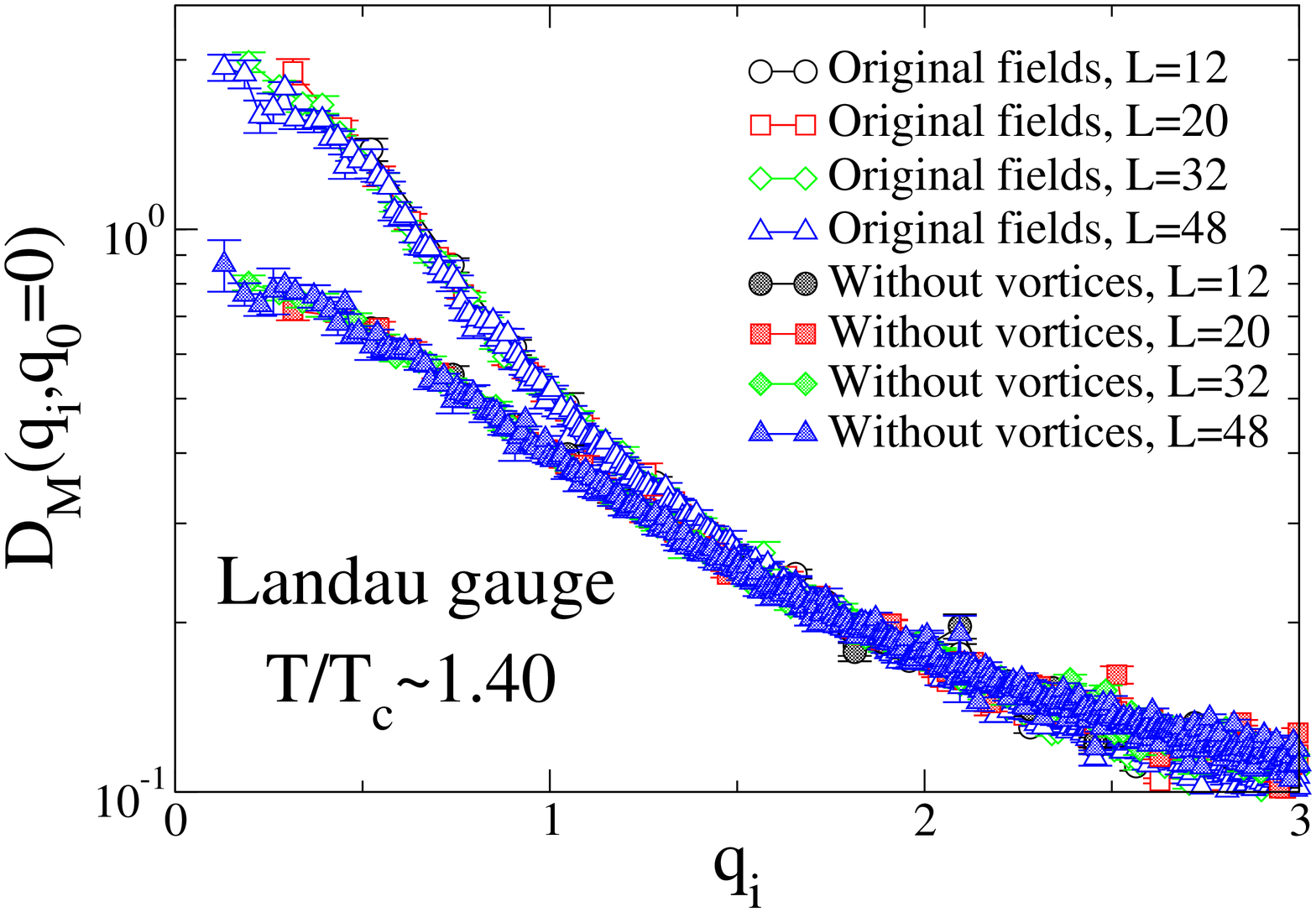}
\caption{The vortex removal: effects of high temperatures (left) and
large lattices (right).}\label{fig3}
\end{figure}

In Fig. \ref{fig3}(left) we show the magnetic and electric propagators at high temperature, $T \approx 6.0 \, T_c$.
There is almost no effect of the vortex removal on the electric propagator. However, as expected, the magnetic propagator is affected by
the vortices  in the infrared region: the removal of the vortices leads to the suppression of gluons with low momenta.
The value of the propagator of
the magnetic gluons (calculated either with or without vortices) in the infrared region is larger compared to the propagator of the
electric gluons in the same region.

\subsection{Volume and Gribov copy effects}

In general, the propagators at the low momentum region can be influenced by the finite volume corrections.
We checked the stability of our results against the variation of the volume of the system in Fig.~\ref{fig3}(right).
Since the data taken at different volumes
follow the same curves within error bars, we conclude that
both original and vortex-removed configurations are independent of the volume.

\begin{figure}[htbp]
\includegraphics[scale=0.29]{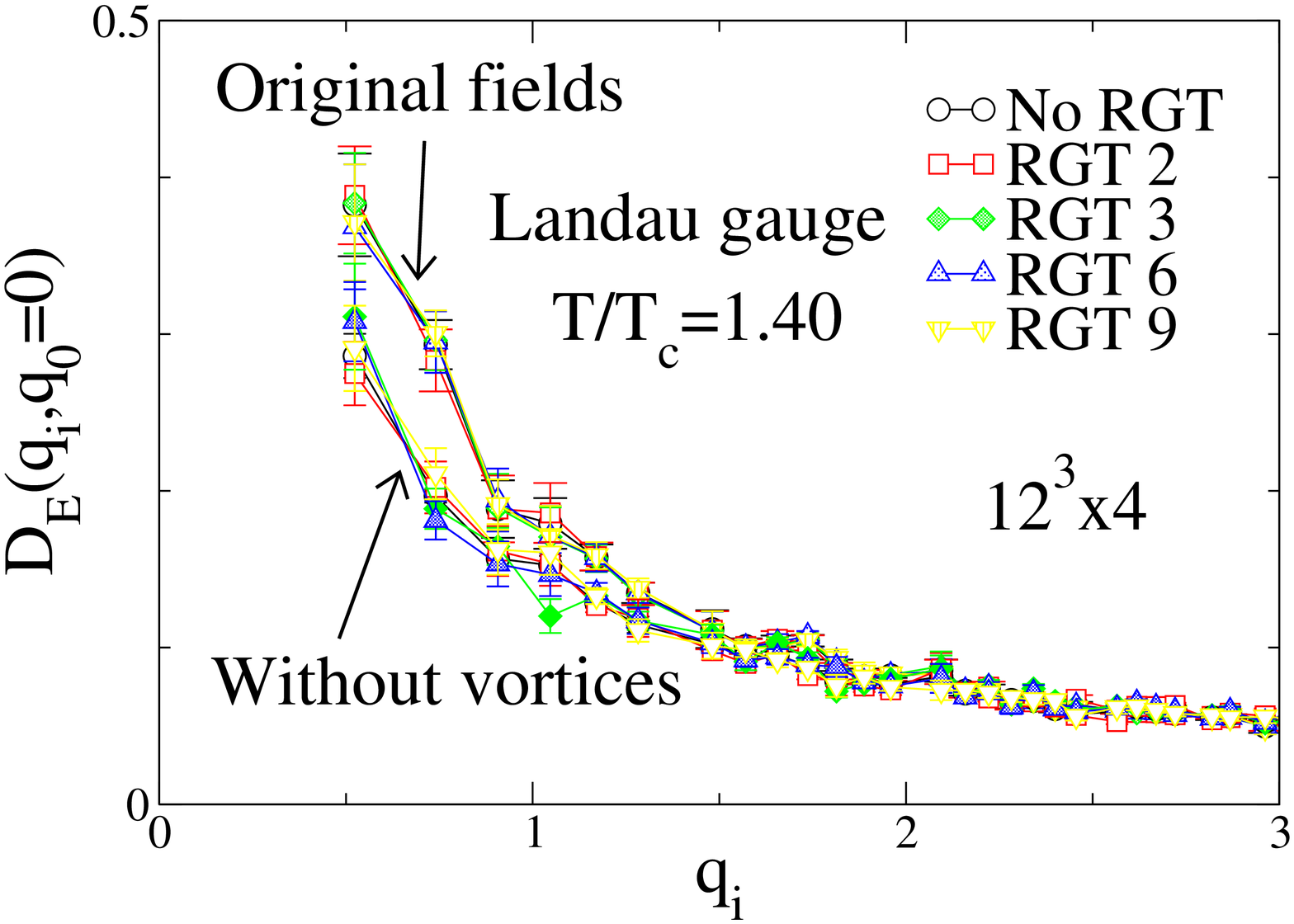}
\hskip -9mm
\includegraphics[scale=0.29]{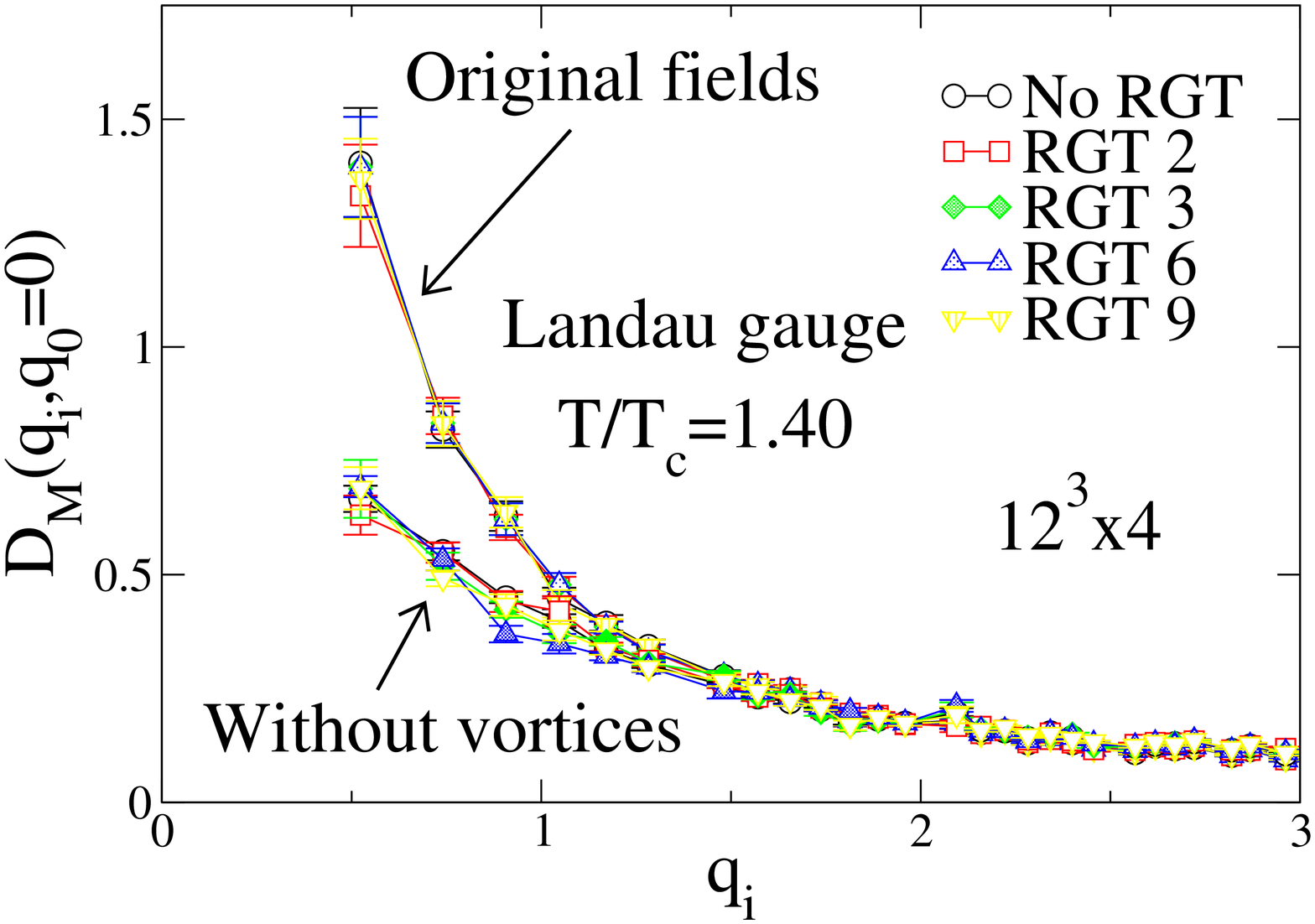}
\caption{A check of the MCG-related Gribov copy effects on the electric (left)
and magnetic (right) gluon propagators in a small volume.}
\label{fig4}
\end{figure}
Another source of error may come from the Gribov copy problem: the global minimization of the functional~(\ref{mcpe})
cannot be achieved exactly using the numerical methods, while the gluon propagator may be affected by the choice of
the minimum. In Fig.~\ref{fig4} we check the robustness of our results against the number of the Gribov copies
given by the number of the starting configurations generated by
random gauge transformations (RGT). One can see that
the {\it qualitative} behavior of the propagators is not affected by the number of the Gribov copies neither for original
nor for the vortex-removed configurations.

\section{Summary}

We found that the presence of the magnetic vortices enhances the propagator
of the magnetic gluons in the low-momentum region at finite-temperature. Alternatively, one can formulate this statement as follows: the vortex
removal procedure suppresses the magnetic propagator at low momenta.
This effect is found both near the phase transition at $T = 1.40 \, T_c$ and in the deep quark-gluon plasma phase at $T = 6.0 \, T_c$.
The propagation of the electric gluon stays almost unaffected by the vortices (the weak effect of vortices is visible close to the critical
temperature and is not seen at the high temperature). We checked that influence of the vortices on the propagation of the
hot gluons is not shadowed by the finite volume or the Gribov copies effects.

In Ref.~\cite{ref:monopoles} the magnetic vortices were suggested to form a light (i.e., low-mass) component of the gluon plasma,
and therefore their removal should naturally suppress the long-distance propagation of the magnetic gluons. Our numerical results
demonstrate that the center vortices are indeed responsible for important nonperturbative properties of the magnetic component of
the quark-gluon plasma.

\end{document}